\documentclass[letter]{aa} 
\usepackage{graphicx}
\usepackage{txfonts}
\usepackage{natbib}
\usepackage{color}
\def\gr{$\gamma$-ray}
\def\FGL{1GFL J0319.7+4130}
\def\IC{IC 310}
\begin{document}
   \title{Very high-energy $\gamma$-ray emission from IC 310}

    \author{A. Neronov
          \inst{1},
          D.Semikoz\inst{2,3}
          \and
          Ie.Vovk
          \inst{1}
          }

   \institute{ISDC Data Centre for Astrophysics, Ch. d'Ecogia 16, 1290, Versoix, Switzerland 
         \and
             APC, 10 rue Alice Domon et Leonie Duquet, F-75205 Paris Cedex 13, France \and
Institute for Nuclear Research RAS, 60th October Anniversary prosp. 7a, Moscow, 117312, Russia
\\
             }


 
  \abstract
   {We search for persistent extragalactic sources of \gr s with energies above 100 GeV with the \textit{Fermi} telescope.}
   {We construct a systematic survey of the extragalactic \gr\ sky at energies above 100 GeV. Such a survey has  not been done before by the ground-based Cherenkov \gr\ telescopes, which have, contrary to \textit{Fermi}, a narrow field of view.}
   {We study a map of arrival directions of the highest energy photons detected by \textit{Fermi} at Galactic latitudes $|b|>10^\circ$ and search for significant point-source-like excesses above the diffuse Galactic and extragalactic \gr\ backgrounds. We identify eight significant point-source-like excesses in this map. }
   { Seven of the eight sources are known TeV blazars. The previously unknown source is identified with the galaxy IC 310, which is situated in Perseus  cluster of galaxies. The source is detected with a significance $6\sigma$ above 30 GeV. We identify two possible scenarii for \gr\ emission from this source. One possibility is  that emission originates from the base of relativistic outflow from the active nucleus, as in the BL Lacs and FR I type radio galaxies. Otherwise \gr\ photons could be produced at the bow shock that is formed as a result of the fast motion of the galaxy through the  intracluster medium. The two models could be distinguished through studying  of the \gr\ signal variability.     }
   {}


   \maketitle
%

   \begin{figure*}
   \begin{center}
  \includegraphics[width=\linewidth]{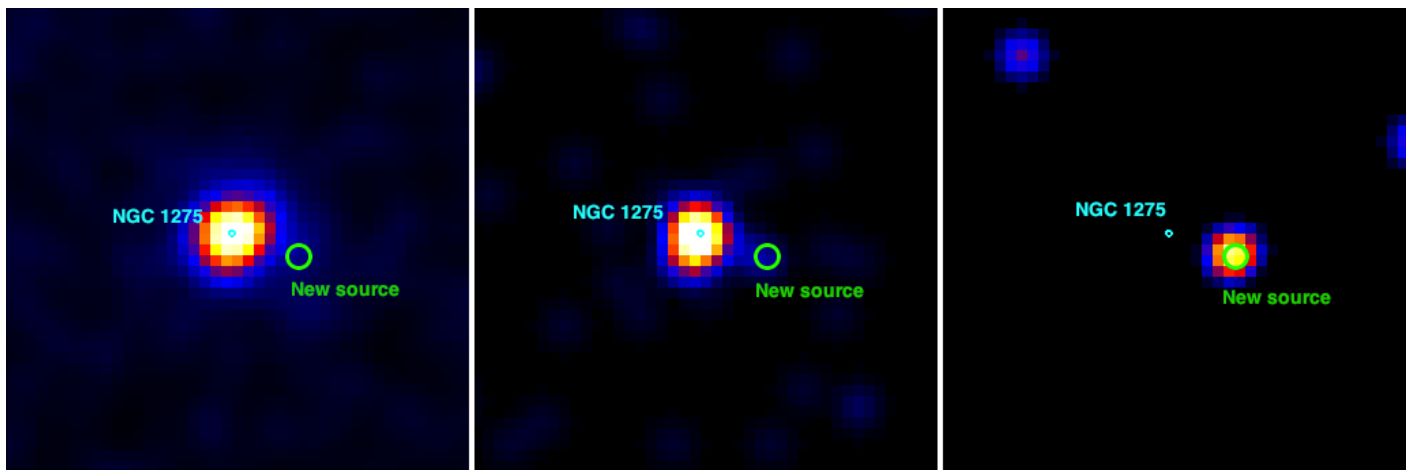}
  \end{center}
   \caption{{\it Fermi} count maps of the sky region around the center of Perseus cluster in 1-10 GeV (left), 10-100 GeV (center) and 100-300 GeV (right) energy bands. The position of {\it Fermi} source \FGL\ is marked by the ellipse corresponding the uncertainty of the source position (95\% confidence level). The green circle of the radius $0.1^\circ$ marks the position of additional source detected at the energies above 100 GeV.   The color scale is linear, from 0 (black) to the maximal number of counts (white).}
              \label{fig:1-10-100GeV}%
              \vskip-0.5cm
    \end{figure*}

\textit{Introduction.}
The number of known astronomical sources of  very-high-energy (VHE) \gr s grew ten-fold over the last five years\footnote{\bf  See e.g. {\tt http://http://tevcat.uchicago.edu/, http://www.mppmu.mpg.de/~rwagner/sources/}}. A large part of the newly discovered sources lie in the Galaxy and were revealed via a systematic scan of the inner Galactic Plane by the HESS telescope \citep{HESS_survey_science,HESS_survey}. The HESS survey has covered an area 0.1 sr in a strip $|l|<30^\circ$, $|b|<3^\circ$.  This covers less than 1\% of the sky. Surveys of  larger regions on the VHE \gr\ sky with the existing ground based Cherenkov \gr\ telescopes are difficult because  the size of the field of view is  too narrow ($5^\circ$ for HESS, $3.5^\circ$ VERITAS telescopes and $3^\circ$ for MAGIC telescope). A previous survey of the northern hemisphere using the Cherenkov telescope Whipple has resulted only in derivation of upper limits on the flux of persistent VHE \gr\ sources \citep{whipple_survey}. The wide field of view MILAGRO \citep{milagro} and Tibet \citep{tibet} arrays have produced a systematic survey of the VHE \gr\ sky. However, the energy threshold of the air shower arrays like MILAGRO and TIBET is  rather high (in the multi-TeV band) so that only sources with spectra extending well above 1 TeV could be detected. 

Contrary to the ground-based Cherenkov \gr\ telescopes, {\it Fermi} has a wide field-of-view and continuously surveys the whole sky on a timescale of $3.2$~hr. Over the first year of operation {\it Fermi} has detected some $1.5\times 10^{3}$ Galactic and extragalactic sources of \gr s with energies above 1~GeV \citep{fermi_catalog}. The smaller collection area of {\it Fermi} ($\sim 1$~m$^2$, compared to $\sim 10^5$~m$^2$ for the ground-based \gr\ telescopes) prevents an extension of the all-sky monitoring with {\it Fermi} to the VHE \gr\ band.

However, the collection area of {\it Fermi} is still sufficient for detecting the brightest \gr\ sources at the energies above 100~GeV. The power of the all-sky monitoring capabilities of {\it Fermi} at the highest energies was clearly demonstrated by discoveries of new VHE \gr\ sources motivated by {\it Fermi} detections of these sources above 10~GeV \citep{ATEL2260,ATEL2486}. The all-sky survey capabilities of space-based \gr\ telescope EGRET at the energies above 10~GeV were used for the search of new VHE \gr\ blazars by \citet{dingus01,10GeV_EGRET} via cross-correlation of arrival directions of highest energy EGRET photons with positions of known sources. 

Below we use {\it Fermi} data  to produce a survey of extragalactic sky at the energies above 100 GeV, i.e. in the energy range accessible for the ground-based \gr\ telescopes. We find that most of the sources visible with {\it Fermi} at the energies above 100~GeV are known TeV blazars. The only source which has not previously been reported as a VHE \gr\ source turns out to be IC 310, which is a head-tail radio galaxy \citep{sijbring98} with possibly a BL Lac type nucleus \citep{rector}. 

Two radio galaxies have previously been reported to be the sources of
\gr s with energies above 100 GeV: M87 \citep{m87,m87_magic,m87_veritas} and Cen A \citep{cena}. These two sources are the two closest Fanaroff-Riley type I (FR I) radio galaxies. The FR I radio galaxies form the "parent" population of BL Lac type blazars \citep{urry95}. They are expected to be weak VHE \gr\ emitters, because the \gr\ flux from these sources is not boosted by the relativistic Doppler effect. In this respect  it  is not surprising that only the two nearest FR I radio galaxies have been seen in the VHE \gr\ band so far.  Both Cen A and M87 are too weak to be detected at 100 GeV in the 1.5~yr exposure of {\it Fermi}. 

\IC\ is situated in Perseus galaxy cluster at the distance of  80 Mpc, which is a factor of 22 and 5 larger than the distances of Cen A and M 87, respectively. \IC\ is, therefore, by 1-2  orders of magnitude more luminous  than that of Cen A and M87.  Besides, \IC\ is not classified as a FR I type radio galaxy. Instead, it is a head-tail radio galaxy \citep{sijbring98}, the type of galaxies usually found in galaxy clusters. It possesses an extended "tail" of the angular size $\sim 15'$  aligned along the direction connecting the center of Perseus cluster and the galaxy. This tail is believed to be produced as a result of the fast motion of the galaxy through the intracluster medium \citep{sarazin88}.
The detection of \IC\ in the VHE band looks less surprising if the hypothesis that the source hosts a low-luminosity BL Lac is adopted \citep{rector}. In this case the mechanisms of VHE \gr\ emission from the source could be the same as in the majority of extragalactic VHE \gr\ sources. 

\textit{Survey of extragalactic sky above 100 GeV with {\it Fermi}.}
We used the data of Large Area Telescope (LAT) on board the {\it Fermi} satellite to generate an all-sky map of arrival directions of \gr s with energies above 100 GeV. The entire list of photons detected by {\it Fermi} in this energy range included 4145 events, collected over the period  from 2008 August 4 till 2010 February 14. A substantial fraction of the events is produced by emission from diffuse and point sources along the plane of the Galaxy. In our analysis of extragalactic sources we exclude the region $|b|<10^\circ$ from the point source search. This leaves 2603 photons available for the analysis. 

To identify significant point sources we first selected clusters of three and more events within circles of the radius $0.1^\circ$, corresponding to the 68\% containment radius of the LAT point spread function at  energies above 100 GeV \footnote{{\tt http://www-glast.slac.stanford.edu/}\\ {\tt software/IS/glast\_lat\_performance.htm}}. For each cluster we calculated the probability to find the observed number of events within the angular resolution circle by chance. To do this, we calculated the overall number $N_{10}$ of events within a circle of 10 degrees radius around the cluster. The chance probability to find a second event within a $0.1^\circ$ around the first event is $p=(1-\cos(0.1^\circ))/(1-\cos(10^\circ))\simeq 10^{-4}$. The chance probability to find a cluster of  at least $K_{0.1}$ events within a circle of the radius $0.1^\circ$  is then 
$P(N_{10},K_{0.1})=\sum_{k=K_{0.1}-1}^\infty \left. p^{k}(N_{10}-1)!\right/(N_{10}-k-1)!k!$

The chance probability $P_{100-300}$ of finding clusters of $K_{0.1}\ge 3$ events in the 100-300~GeV energy band  is less than $10^{-5}$ everywhere in the extragalactic sky. The list of $K_{0.1}\ge 3$ excesses found at Galactic latitudes $|b|>10^\circ$ is given in Table \ref{tab:catalog}. In the same table we also give the number of \gr s within the circle of the radius $0.2^\circ$, which corresponds to the 95\% containment radius of LAT PSF above 100~GeV. In principle, the estimate of the chance probability of finding a cluster of $\ge 3$ photons in the background events could be improved if known {\it Fermi} sources are removed from the background. However, the only $< 10^2$ photons with energies above 100~GeV could be associated to the known {\it Fermi} sources \citep{neronov10}, which means that the correction introduced by  "masking" the known sources in the background estimates would be small.

All but one detected clusters of $K_{0.1}\ge 3$ events listed in Table \ref{tab:catalog} could be identified with the known TeV blazars. The only source that does not belong to the blazar class and was not previously detected in the VHE energy band is the source \#2 at $RA=49.14\pm 0.08, \ DEC=41.30\pm 0.06$.

\begin{table}
\begin{tabular}{lllllllll}
\hline
&Name & RA & DEC & $K_{0.1}$ & $K_{0.2}$ & $P_{100-300}$ \\
\hline
1&3C 66A & 35.67 & 43.06 & 4 & 5 & $<5\times 10^{-7}$  \\
{\bf 2}&{\bf IC 310} & {\bf 49.14} &  {\bf 41.30} &   {\bf   3   } & {\bf 3} & {\bf $ 6\times 10^{-6}$}\\
3&1ES 0502+675 & 77.15 &   67.61 &   3 &  3 & $7\times 10^{-6}$\\     
4&Mrk 421 & 166.11 & 38.20 &  20  & 24 & $<5\times 10^{-7}$\\  
5&PG 1553+113 & 238.90 & 11.24  &    3 & 5 &$<5\times 10^{-7}$ \\
6&Mrk 501 & 253.49 & 39.77   &  10  & 10 & $<5\times 10^{-7}$\\ 
7&PKS 2005-489 & 302.37 & -48.85 &    4 & 4 & $<5\times 10^{-7}$  \\ 
8&PKS 2155-304 & 329.74 & -30.23 &  8 & 9 & $<5\times 10^{-7}$     \\
\hline
\end{tabular}
\caption{Catalog of \gr\ sources detected by LAT at  energies above 100 GeV at $|b|>10^\circ$.}
\label{tab:catalog}
\end{table}

\textit{VHE \gr\ emission from IC 310.}
The excess at the position of  source \#2 has three photons at  energies above 100~GeV. The chance probability that such an excess is due to a chance coincidence is  ruled out with probability $(1-P_{100-300})\simeq 99.9994\%$, which corresponds to a $4.5\sigma$ significance of the source detection above 100~GeV.

{\it Fermi} images of the sky region around source \#2 are shown in Fig. \ref{fig:1-10-100GeV}. The new source is situated approximately $0.6^\circ$ to the southwest from the bright {\it Fermi} source 
1FGL J0319.7+4130, which is identified with the radio galaxy NGC 1275 in the center of Perseus galaxy cluster. The spectrum of the new source is harder than the spectrum of NGC 1275, so that emission from NGC 1275 dominates in the 1-10 GeV energy band, while emission from the new source dominates the signal above 100~GeV. 

At energies above 30~GeV, 95\% of events from a point source are contained in a circle of the radius $0.3^\circ$. This means that events from the bright source  1FGL J0319.7+4130 do not affect the signal from the newly detected source $0.6^\circ$ away. Taking this into account, we included events with energies above 30~GeV in the analysis. We found two  more events from the source within the $0.1^\circ$ search circle in the 30-100~GeV band. To calculate the probability $P_{30-100}$ of finding two additional events at the source position  one can use modification of  the expression for the 100-300 GeV band with a substitution $K_{0.1}-1\rightarrow K_{0.1}$ and $N_{10}-1\rightarrow N_{10}$.  Doing this we find  $P_{30-100}=1.1\times 10^{-4}$.   The probability to find $\ge 2$ additional photons in the 30-100~GeV to a cluster of  $\ge 3$ photons in the 100-300~GeV band is  $P_{100-300}P_{30-100}\simeq 6.6\times 10^{-10}$ (which corresponds to $6\sigma$ for Gaussian statistics). Equivalently, the combined probability to find a 3.9 and 4.5$\sigma$ excess  in the two bands is $\simeq 2\times 10^{-8}$, which corresponds to the significance of source detection at the level of $5.6\sigma$ (integrating the probability density outside constant likelihood contour for Gaussian statistics). 

In order to verify the detection of the source with {\it Fermi}, we  performed a standard Fermi analysis\footnote{{\tt http://fermi.gsfc.nasa.gov/ssc/data/}\\ {\tt analysis/scitools/}} for the sky region containing \IC. In this analysis we used photons with energies between 10 GeV and 300 GeV, collected from a circular area with a radius of 5 degrees around the position of \IC. We included all sources mentioned in the Fermi first year catalog \citep{fermi_catalog} for this region of the sky. All the sources had freely varying  normalizations and spectral indices. We assumed that the spectra of all the sources included in the likelihood analysis had power-law shape. 

\begin{figure}
\includegraphics[width=0.7\linewidth]{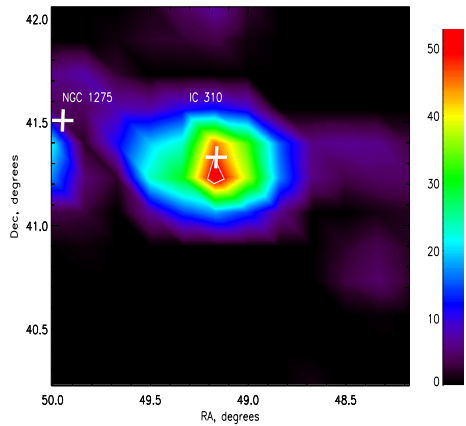}
\caption{Map of the  TS values for an additional source near NGC 1275 at energies above 10~GeV. The white contour corresponds to the 95\% uncertainty of the source position.}
\label{fig:TSmap}
\end{figure}

The likelihood analysis has resulted in the detection of \IC\ with the Test Statistic (TS) \citep{mattox}  value of 53, which corresponds to approximately $7\sigma$ source detection significance and  is compatible with the $6\sigma$ source detection significance found from the direct photon counting above 30 GeV. Figure \ref{fig:TSmap} shows the map of TS values generated with {\it gtlike} tool by varying the position of the source on the grid of 12 by 12 positions with a 0.1 degree step. The uncertainty of the source position found from the TS map is also compatible with the one found from the direct photon counting. 

\begin{figure*}
   \begin{center}
  \includegraphics[width=\linewidth]{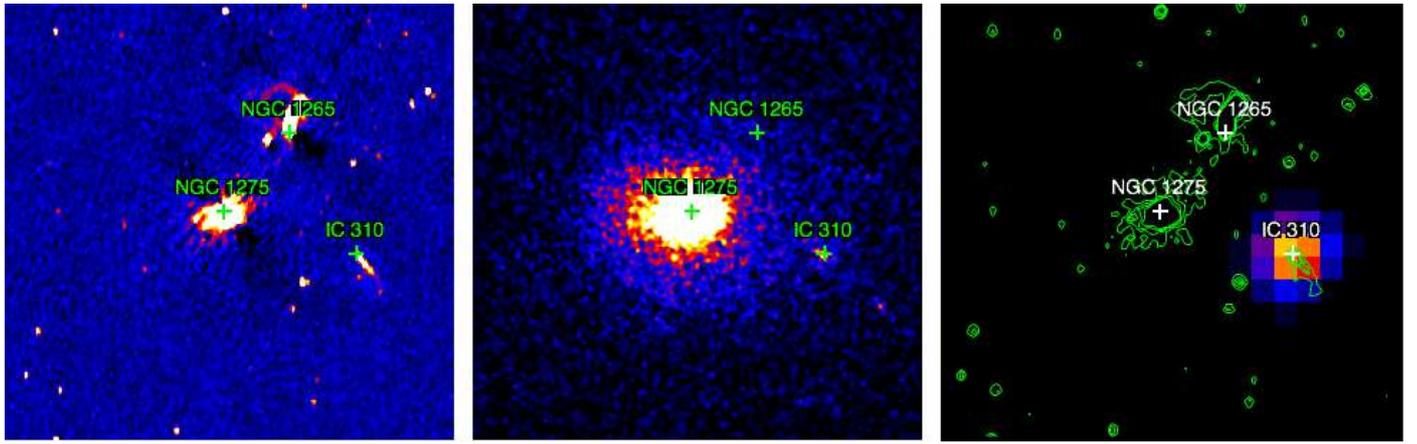}
  \end{center}
   \caption{Image of Perseus galaxy cluster in radio band from WENSS sky survey (left), in the X-ray band from the {\it ROSAT} all sky survey (middle) and the  {\it Fermi} image in the 100-300 GeV energy band (right). The green contours on the right panel correspond to the radio image from the left panel. }
              \label{fig:ROSAT+radio}%
    \end{figure*}

Figure \ref{fig:ROSAT+radio} shows a comparison of the Fermi image above 100~GeV with the images of the same region of the sky in X-ray ({\it ROSAT} all sky survey\footnote{\tt http://www.xray.mpe.mpg.de/cgi-bin/rosat/rosat-survey}) and radio (WENSS survey \citep{wenss})  bands obtained through the SkyView interface\footnote{\tt http://skyview.gsfc.nasa.gov/}. The radio and X-ray source at the position of the new \gr\ source is the galaxy IC 310, which is a part of Perseus cluster. There are no other bright  radio or X-ray sources within the $0.1^\circ$ degree circle around the position of the \gr\ source. This provides an unambiguous identification of the source with IC 310.
 
\begin{figure}
\includegraphics[width=\linewidth]{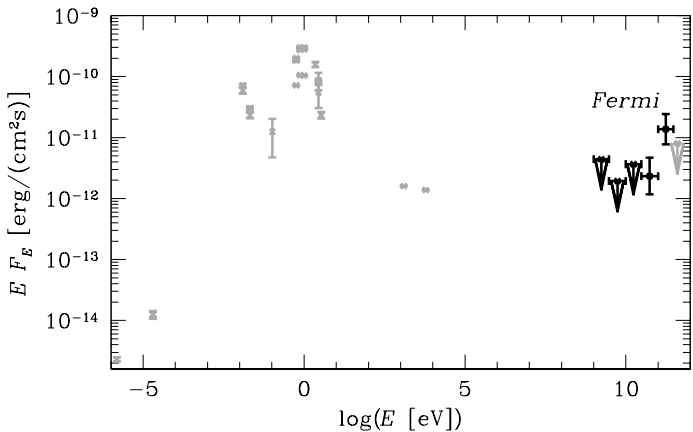}
\caption{Spectral energy distribution of \IC. Grey points are taken from NED. Black points are {\it Fermi} measurements.}
\label{fig:SED}
\vskip-0.3cm
\end{figure}

In order to estimate the flux from the source in the 30-300 GeV energy  range, we calculated the exposure of {\it Fermi/LAT} in the direction of the source in this energy band with the {\it gtexposure} tool. Collecting the photons from the 95\% containment circles of the energy-dependent point spread function for the front and back-converted photons \citep{lat-performance},  we find the estimates of the source flux in two energy bins, 30-100~GeV and 100-300~GeV are $F_{30-100}= 2.3_{-1.2}^{+2.3}\times 10^{-11}$~erg/(cm$^2$s) and $F_{100-300}= 1.4_{-0.6}^{+1.1}\times 10^{-11}$~erg/(cm$^2$s), respectively. These estimates are shown in Fig. \ref{fig:SED} together with the multiwavelength data on IC 310 collected from the NASA Extragalactic Database\footnote{\tt http://nedwww.ipac.caltech.edu/}. 

At energies below 30~GeV, the signal from NGC 1275 could give a non-zero number of photons at the position of the new source. In order to estimate the background at the source position  we adopted the following procedure. We have chosen three "background" circular regions of the radius $0.3^\circ$ situated at the same angular distance from NGC 1275 as the new source, but at the position angles $90^\circ,\ 180^\circ$ and $270^\circ$, with respect to the direction from NGC 1275 toward the new source. The signal of NGC 1275 should give approximately the same number of counts in the three "background" regions and in the "source" region, which we choose to be  a circle of the radius $0.3^\circ$ around the source position.  We found that the signal in the source circle is compatible at $\le 3\sigma$ level with the  signal in the background regions in the energy bins 1-3~GeV, 3-10~GeV and 10-30~GeV, see Fig. \ref{fig:SED}.

\textit{Discussion.}
Known VHE \gr\ loud Active Galactic Nuclei (AGN) are divided into several classes. Most of the sources are BL Lacs, which are relativistically beamed versions of Fanaroff-Riley type I  radio galaxies. Two of the detected sources are the nearest  FR I radio galaxies themselves (M87 and Cen A). One source, 3C 279, belongs to the Flat Spectrum Radio Quasar (FSRQ) class.  Unless \IC\ proves to be a weak BL Lac type object, the detection of VHE \gr\ emission from a head-tail radio galaxy provides a new class of extragalactic VHE \gr\ sources. If the interpretation of the source as a weak BL Lac \citep{rector} is adopted, the source still has unusual properties due to its unusual for BL Lacs radio morphology and host galaxy (lenticular, instead of a giant elliptical) \citep{nilson}.

In FR I radio galaxies and BL Lacs the VHE \gr\ emission is most probably produced in the innermost part of the jet that is ejected by the supermassive black hole. For IC 310, it is not clear a priori if the \gr\ emission is powered by the same mechanism. An alternative possibility is that \gr s are produced at the bow shock formed in interaction of relativistic outflow from the fast moving galaxy with the intracluster medium. In this respect, the \IC\ -- Perseus cluster system might be similar to a much smaller scale PSR B1259-63 system, in which \gr\ emission is produced at the bow shock formed in interaction of relativistic outflow from a pulsar  moving through a dense wind of a companion star \citep{tavani97}. 
 
Angular resolution of \gr\ telescopes is only marginally  sufficient to resolve the bow shock surface in the IC 310 -- Perseus cluster system. The angular length of the "tail" of the bow shock visible in the radio band is $\sim 15'$. The uncertainty of  the {\it Fermi} source position is  $\sim 4'$, which is smaller than the length of the tail. The {\it Fermi} data indicate that most of the \gr\ emission is produced in the "head" part of the source (see Fig. \ref{fig:ROSAT+radio}). At the same time, the {\it Fermi} angular resolution is not sufficient to distinguish between emission from the "head" of the bow shock and the emission from the base of the jet near the supermassive black hole, which powers the  source activity.

 The crucial test, which would allow to distinguish between the two mechanisms would be the (non) detection of variability of \gr\ emission from the source. Indeed, in the BL Lac  type models the  \gr\ emission is expected to be variable at different timescales, down to the timescale of light-crossing of the central supermassive black hole. On the other hand, if the observed \gr\ emission is produced at the bow-shaped contract surface between the AGN outflow and the intracluster medium, the \gr\ source has $\sim$kpc scale size. This means that the \gr\ emission could not be variable on  timescales much shorter than $\sim 10^3$~yr.

Variability of the \gr\ signal from \IC\ could not be studied with {\it Fermi}, which has detected only five \gr s from the source at energies above 30~GeV. We verified that the five detected events did not come within a narrow time window, which would indicate the possibility of a strong flare from the source. 

The presence or absence of variability of the VHE \gr\ emission from \IC\ could be readily verified in observations with ground based \gr\ telescopes. A previous observation of the region around Perseus cluster with Whipple telescope has resulted in an upper limit one the source  flux  \citep{perkins06}. However, this upper limit is comparable to the {\it Fermi} measurement of the source flux, so that no conclusion about the presence or absence of long-term variability of the source could be drawn from the comparison of Whipple and {\it Fermi} observations.  It is clear that observations of the source with the new generation of ground-based \gr\ telescopes VERITAS or MAGIC (the source is in the northern hemisphere, not easily accessible for HESS) would give much higher signal statistic at energies above 100~GeV, so that the hypothesis of the flux variability could be easily tested\footnote{Following the  discovery of the source above 100 GeV with {\it Fermi}, the detection in the same energy band was confirmed by the  MAGIC telescope \citep{MAGIC_ATEL}.}.

From Fig. \ref{fig:ROSAT+radio} is is immediately clear that not every head-tail radio galaxy in the Perseus cluster emits in the VHE \gr\ band  at the {\it Fermi} sensitivity level. The image of the cluster, shown in this figure, includes another prototypical head-tail galaxy, NGC 1265. This source is clearly identified in the radio band, but, contrary to  \IC, does not show significant X-ray and VHE \gr\ emission. A comparison of the physical parameters of \IC\ and NGC 1265 (e.g. velocity through the intracluster medium, overall power of relativistic outflow etc.) could help to clarify the conditions under which particle acceleration and VHE \gr\ emission in this type of sources occurs.

\begin{acknowledgements}
This work was supported by the Swiss National Science Foundation grant PP00P2\_123426/1. We thank L.Foschini for discussions of the subject.

\end{acknowledgements}

\end{document}